\documentclass[aps,prb,twocolumn,superscriptaddress]{revtex4-1}

\usepackage{graphicx}

\begin{document}

\title{Electronic Structures of Black Phosphorus Studied by Angle-resolved Photoemission Spectroscopy }

\author{C. Q. Han}
\author{M. Y. Yao}
\author{X. X. Bai}
\author{Lin Miao}
\author{Fengfeng Zhu}
\author{D. D. Guan}
\author{Shun Wang}
\author{C. L. Gao}
\author{Canhua Liu}
\author{Dong Qian}
\email{dqian@sjtu.edu.cn}
\author{Y. Liu}
\author{Jin-feng Jia}
\affiliation{Key Laboratory of Artificial Structures and Quantum Control (Ministry of Education), Department of Physics and Astronomy, Shanghai Jiao Tong University, Shanghai 200240, China}

\date{\today}

\begin{abstract}
Electronic structures of single crystalline black phosphorus were studied by state-of-art angle-resolved photoemission spectroscopy. Through high resolution photon energy dependence measurements, the band dispersions along out-of-plane and in-plane directions are experimentally determined. The electrons were found to be more localized in the ab-plane than that is predicted in calculations. Beside the k$_z$-dispersive bulk bands, resonant surface state is also observed in the momentum space. Our finds strongly suggest that more details need to be considered to fully understand the electronic properties of black phosphorus theoretically.
\end{abstract}


\maketitle

Since the experimental realization of real two dimensional (2D) material -- graphene\cite{grahene}, great efforts have been devoted to study similar 2D semiconducting systems that are believed to be very useful for future applications\cite{2Dmaterials,2D}. As a layer-structured and narrow-gap elemental semiconductor with a direct energy gap of about 0.33 eV, black phosphorus (BP) has received more and more attention in recent years\cite{park1,park2,Zhang,Sun,NIlges,Marino}. Similar to carbon, phosphorus exists in a number of allotropic forms. Orthorhombic BP is the most stable form of phosphorus under normal conditions. BP has a puckered layer structure (Fig. 1a) that has a honeycomb network similar to a graphene layer\cite{structure1,structure2}. The investigations of bulk BP showed many interesting physical properties. BP can server as the electrode material for Lithium-Ion batteries\cite{park1,park2,Sun,Stan,Marino,Nagao}. Structural phase transition from orthorhombic structure to the rhombohedral structure and to the simple cubic structure was found under pressure accompanied by semiconductor-semimetal-metal transition\cite{phasetransition1,phasetransition2,phasetransition3,phasetransition4,phasetransition5}. Furthermore, it has been reported that the BP single crystal shows superconductivity with Tc higher than 10K under high pressure\cite{SC1,SC2}. Recently, mechanical peel-off method was used to get nanometer thick BP layers that were successfully fabricated to a transistor\cite{Yuanbo}. Single layer BP was proposed to be a direct gap semiconductor\cite{singlelayer1,singlelayer2}, though it has not been obtained experimentally. In spite of its various interesting properties as mentioned above, so far, the electronic structures of BP are not completely understood yet. Asahina et al. have calculated the band structure of BP based on the tight binding method, which showed that BP had a direct energy gap of about 0.3 eV at the Z point\cite{cal1}. Goodman et al. also studied the electronic structure of BP using local orbital method and got the similar result\cite{cal2}. Using \textit{ab initio} calculation, Y. Du et al. found that the energy bands wer more complex and the band gap was at the $\Gamma$ point\cite{cal3}. Experimentally, though normal emission photoemission with limited resolution had been carried out to measure the energy bands along the k$_z$ directions and energy gap was determined to be at the Z point \cite{photoemission1, photoemission2}, high resolution detailed band structures in the whole Brillouin zone (BZ) have not been reported. In this work, by carefully tuning the incident photon energy, we experimentally determined the band dispersions along several high symmetry directions in momentum space using high resolution ARPES on single crystalline BP. Comparing with the reported calculations, we found that the valence band near the Fermi level splits into two bands resulting in strong suppression in bandwidth along k$_z$ direction, while the bandwidth in the plane (ac plane, as shown in Fig. 1(a)) becomes larger. Beside the bulk energy bands, surface resonant states were also observed in the momentum space.

BP was synthesized under high pressure and high temperature conditions using white and red P separately as starting materials, as previously reported\cite{Sun,Endo}. The temperature dependence of the bulk resistivity and Hall effect measurements of the samples show that the BP we studies are p-type semiconductors (present elsewhere). All the samples were cleaved at 30 K resulting in well-ordered and shinning surfaces (ac plane). The ARPES measurements were performed using 70 - 130 eV photons at Advanced Light Source beamlines 4.0.3 using Scienta R4000 analyzers with base pressures better than 5$\times$10$^{-11}$ torr. Energy resolution was better than 15 meV and angular resolution was better than 0.02 \AA$^{-1}$. Different polarization lights were used to reduce the matrix element effect in ARPES measurement\cite{zxshen}. The position of the Fermi level was referenced to a copper plate in electrical contact with the samples. No charging effect was observed during measurements at low temperatures.

\begin{figure*}
\centering
\includegraphics[width=13cm]{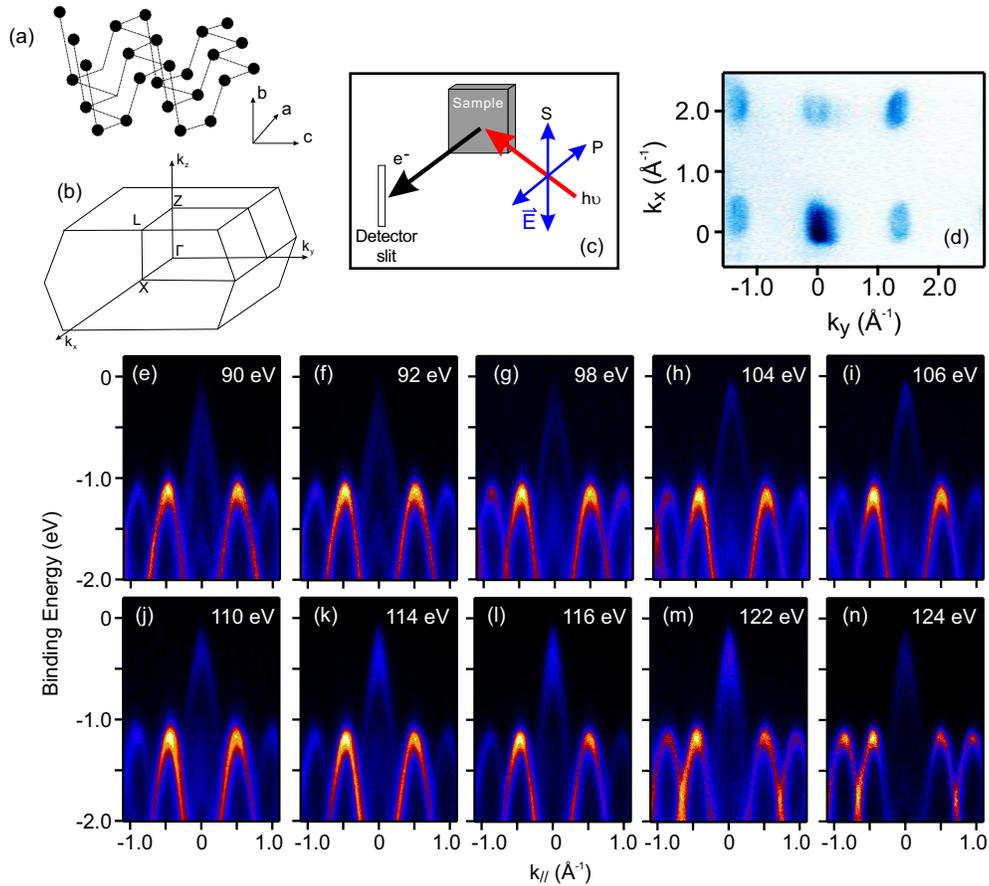}
\caption{(a) Sketch of the BP crystal structure. (b)Brillouin zone of orthorhombic BP. (c)S and P-polarization lights were used in the experiments. (d)Constant energy contour at binding energy of -0.3 eV. Rectangular symmetry is consistent with the bulk orthorhombic crystal structure. (d)-(m) ARPES spectra near normal emission position along $\Gamma(Z) - X(L)$ direction using different incident photon energy.
}
\end{figure*}

Experimentally, band dispersion along k$_z$ direction (b-axis) can be determined by changing the energy of the incident photons. One set of APRES spectra using different photon energy along $\Gamma(Z) - X(L)$ are shown in Figure 1. Close to the Fermi energy, hole-like valence band is centered at k$_{\parallel}$=0 above binding energy of $\sim$ -1 eV. Constant energy contour at binding energy of -0.3 eV (Fig. 1(d)) shows rectangular symmetry, which is consistent with bulk crystal structures. Seen from Fig. 1(e)-(n), the intensity and sharpness of this hole-like band varies with the changing of the photon energy. At each photon energy (from 90 eV to 128 eV with an interval of 0.5 eV), we take the energy distribution curves (EDCs) right at k$_{\parallel}$=0 and make a new plot in Figure 2 to determine the band's k$_z$ dispersion. Fig. 2 (a) and (b) show the image plot and corresponding EDCs of the measured bands along the $\Gamma$-Z-$\Gamma$ direction. A whole BZ along b-axis is covered. Seen from the figure 2(a) and (b), there are two dispersive bands centered at about -0.5 eV and -1.5 eV (Black dotted lines mark the two bands in the spectra.) and some non-dispersive features labeled by green arrow. Based on the periodicity observed on the experiments, we can calculate the momentum position where the valence bands reach the maximum. It turns out that the valence band maximum is at Z point and corresponding photon energy is $\sim$ 104 eV, which is consistent with the previous photoemission results\cite{photoemission1, photoemission2}. The band minimum point is at $\Gamma$ point and corresponding photon energy is $\sim$ 90 eV and 122 eV. From $\Gamma$ to Z point, the topmost valence band disperses from $\sim$ 0.75 eV below Fermi level towards the Fermi level. It turns back at Z point without crossing the Fermi level, which presents the semiconducting nature of BP. Fig. 2(c) shows spectra from BP taken at Z point and from polycrystalline Cu plate as reference. The energy gap between the BP's valence band maximum and the Fermi level is about 0.12 eV (determined by the peak position of the valence band spectra) or 0.05 eV (determined by the leading edge of the valence band spectra) at 30K. Considering the full energy gap between valence band and conduction band of BP is about 0.3 eV, ARPES result is consistent with bulk macroscopic p-type behavior (Fermi level is more close to the valence band). At about 1.5eV below the Fermi energy, there is the second energy band with "M" shape dispersion, which has a local minimum at Z point. The energy gap between upper and lower bands is about 0.5 eV. Beside two dispersive bands, some non-dispersive features are also visible near $\Gamma$ points (indicated by green arrow). This feature is so broad that we think there should be multi-peaks though we can not clearly resolve them experimentally. Similar to bulk valence band, the non-dispersive bands don't cross the Fermi level. Because its non-k$_z$ dispersive character, we think these features should be related to surface states. Under suitable photo energy, one of the surface states was resolved in the ac-plane as discussed below.

\begin{figure}
\centering
\includegraphics[width=8.6cm]{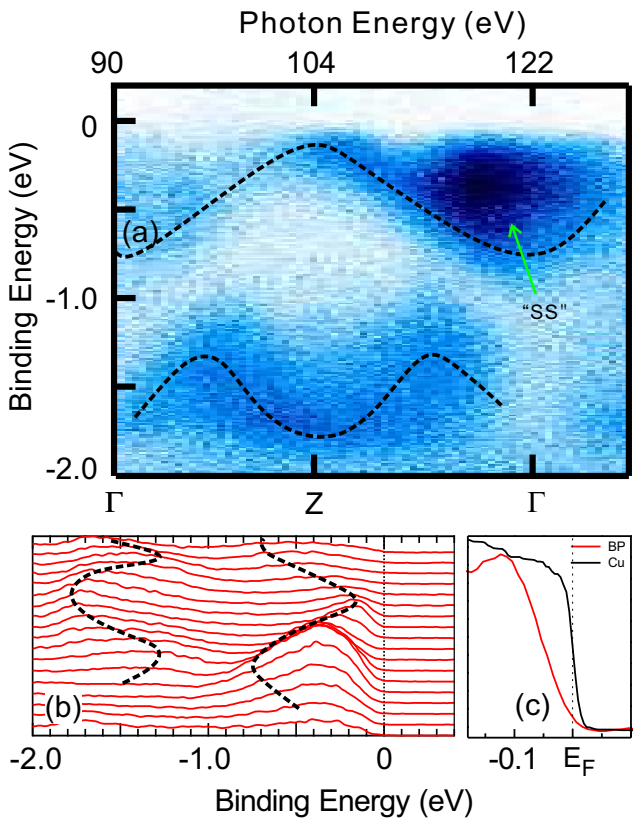}
\caption{(a)Experimental energy band dispersion along the k$_z$ directions and (b) corresponding EDCs . (b)Energy gap between the valence band maximum  and Fermi level. Black curve is the spectrum from Cu as a reference.}
\end{figure}

\begin{figure}
\centering
\includegraphics[width=8.6cm]{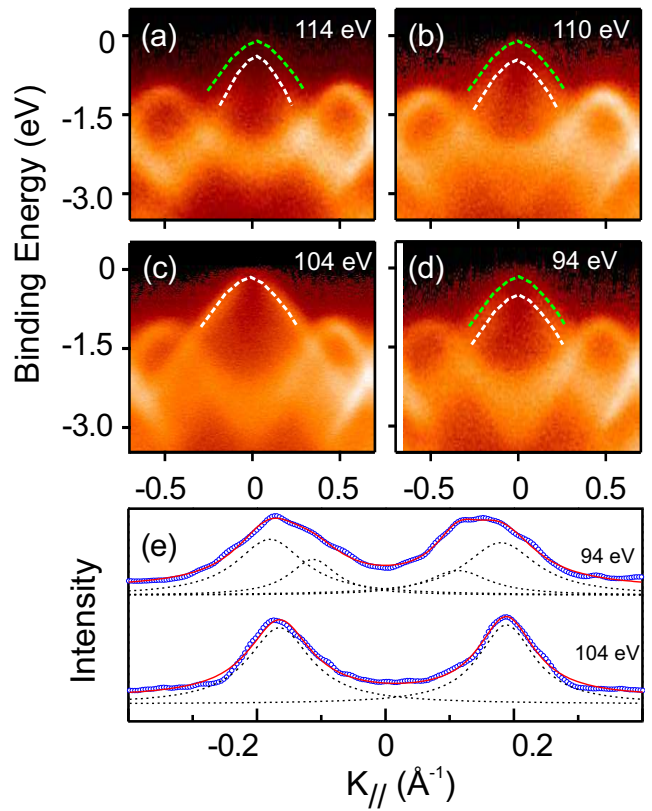}
\caption{High resolution ARPES spectra near $\bar{\Gamma}$ point using different photon energy (a) hv=114 eV. (b)hv=110 eV. (c)hv=104 eV. (d)hv=94 eV. White dashed line is the bulk band. Grean dashed line is the surface band. (e)MDCs at binding energy of -0.75 eV. Open symbols are the experimental data. Solid and dashed lines are the fitting results.}
\end{figure}

Figure 3 presents high resolution ARPES sepctra near $k_{\parallel}=0$ point using different photon energies. Obviously, the hole like valence band is sharpest using 104 eV photon energy. Away from 104 eV, the spectra become weak and broad. Fig. 3(e) shows two momentum distribution curves (MDCs) near k$_{\parallel}=0$ at binding energy of -0.75 eV using 94 eV and 104 eV photo energy, respectively. Under 104 eV photon energy, the MDC can be nicely fitted using two lorentzian peaks. However, under 94 eV photon energy, the MDC shows more than two peaks, which indicates that the broadening of the spectra is due to multi-band effects instead of the increasing of scattering rate. The MDC of 94 eV can be fitted using four lorentzian peaks. Seen from the fitting results, the position of the outer two peaks is nearly the same as the spectra of using 104 eV photon. Only the inner two peaks move to small momentum, so the outer two peaks should from the surface states and the inner peaks should from the bulk states. Know from Fig. 2(a), 104 eV photon energy corresponds to Z point, the valence band reaches the maximum at this momentum position. Away from Z point, valence band moves down to higher binding energy. At Z point, those two bands coincide resulting sharp spectra. When the bulk valence disperses away from the band maximum, the surface states remain. In Fig. 3, white and green dashed lines present the bulk valence band and surface band. This surface state has the same in-plane dispersion relation as that of bulk valence band, indicating it is a resonance surface state.

In further, we mapped the bulk band dispersion in the Z-L-Z ($k_z$=$\pi$) and X-$\Gamma$-X ($k_z$=0) directions. Figure 4(a) and (c) show the ARPES spectra measured using 104 eV photons with different linear polarization along Z-L-Z-L direction. Figure 4(e) shows the ARPES spectra along X-$\Gamma$-X-$\Gamma$ direction. Low energy bands in two in-plane BZs were measured which helps us to determine all the bands because some bands in the first BZ are weak or even un-observable, but can be resolved in the second BZ due to matrix element effects. To further reduce the matrix element effects, p-polarization light is also used as shown in Fig. 4(c). In fact, seen from the Fig. 4, all the bands can be well resolved either by looking at different BZ or using different polarization. In order to enhance the band dispersion, second derivative image (SDI) plots are shown in Fig. 4(b), (d) and (f) correspondingly.

\begin{figure}
\centering
\includegraphics[width=8.6cm]{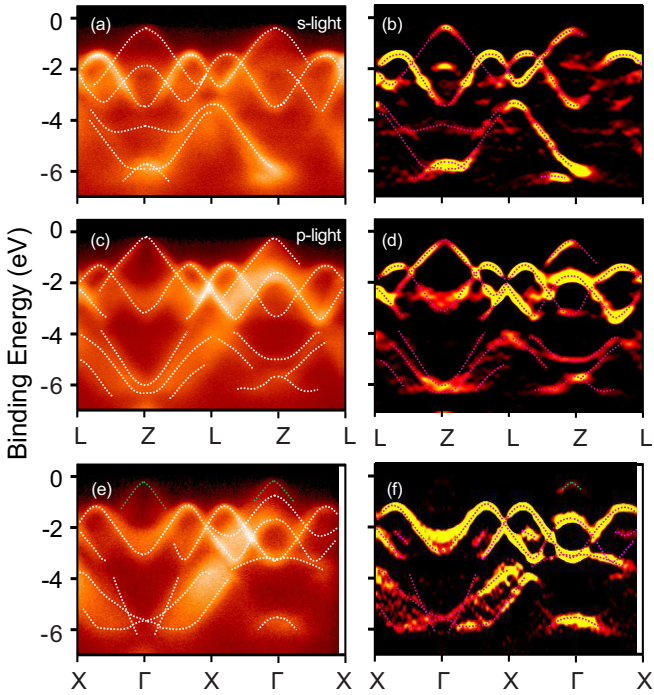}
\caption{The experimental band dispersions in the ac-plane (a)Along L-Z-L-Z direction using s-polarization light. (b)SDI plots of (a). (c)Along L-Z-L-Z direction using p-polarization light. (d)SDI plot of (c). (e) Along X-$\Gamma$-X-$\Gamma$ direction using s-polarization light. (f) SDI plots of (e). White and red dashed line marks the bulk bands. Green dashed line marks the surface bands}
\end{figure}

By tracing the peak position in the ARPES spectra as well as in the SDI plots in Fig. 4, we plot all the bulk bands that were observed in single BZ in figure 5. Previously reported calculation results\cite{Yuanbo,cal1,cal2} were overlaid on the top of experimental results. The red dots are experimental data and blue solid lines are calculated bands. Seen from Fig. 5(a), there are dramatic differences between the experimental bands and the the calculated bands along k$_z$ direction. Close to Fermi level, there is single parabolic like band with a width of about 1.75 eV in calculation. Experimentally, within this energy region ($<$ 2eV), we observed two bands (Band-I and Band-II). The bandwidth of the the topmost band (band-I) is about 0.7 eV, which indicated bulk BP would behave more two dimensionally than theoretical expectation. This effect is consistent with our recent transport measurements (present elsewhere). In further, from the band mapping in the plane, we reveal that the observed band renormalization along k$_z$ direction is caused by the splitting of the calculated band-1. Fig. 5(b) and (c) present the bands in the plane along L-Z-L and X-$\Gamma$-X direction, respectively. In calculations, from Fermi level to binding energy of about -6 eV, there are two set of band complexes, between which there is a energy gap. This gap is smaller in reference [21] (Fig. 5(d)). However experimentally, as shown in Fig. 5(b), we found that these two band complexes overlapped. The total bandwidth in the plane becomes about 6eV that is nearly ten times larger than that along out-of-plane direction. Near the valence band maximum (fig. 5 (b) and (d)), within the experimental uncertainty, experimental band agrees very well with the calculations. Between the binding energy of 0 to -3 eV, there are two bands in calculation (labeled by blue "1" and "2"). However, clearly resolved in experiments, there are three bands (labeled by red "I", "II" and "III"). This discrepancy can be understand based on the splitting of band-1. Seen from the Fig. 5(a), the shapes of the band-I, band-II and band-1 looks similar, so we think the band-1 in calculation actually splits into two bands (band-I and band-II) in real materials. Due to the existence of band-II, band-2 is pushed to high binding energy to form the experimental observed band-III in L-Z-L direction.

\begin{figure}
\centering
\includegraphics[width=8.6cm]{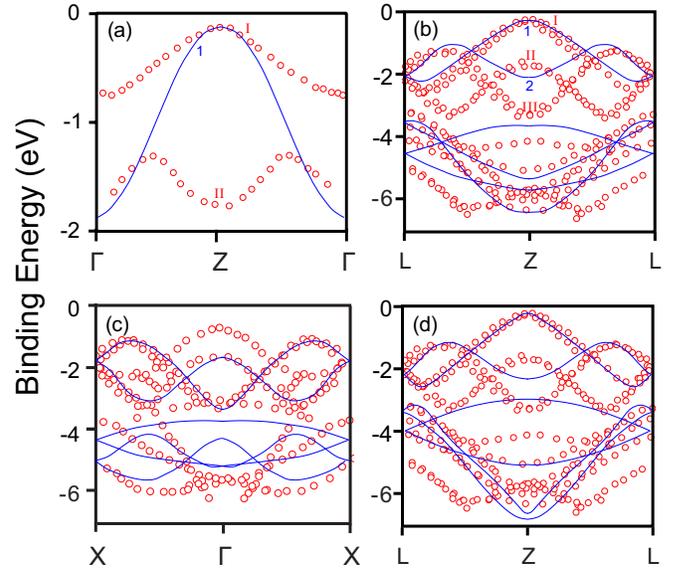}
\caption{Comparison the calculated band dispersion with the experimentally determined band dispersion along (a)$\Gamma$-Z-$\Gamma$ direction. (b)L-Z-L direction. (c)X-$\Gamma$-X direction. The calculation bands is from reference [24]. (d) L-Z-L direction. The calculation bands is from reference [21].}
\end{figure}

In summary, we studied the detailed band dispersions in the ac-plane as well as out-of-plane in BP. Consistent with calculation, the valence band maximum is at Z point. However, the observed band width along k$_z$ direction is much smaller than predicted in calculation, which implies that electrons are more localized in two dimensional plane. Besides the bulk energy bands, resonant surface states are also observed, which may play important roles when the thickness of BP is reduced to several layers. Our findings imply that more details in BP should be considered in order to quantitatively explain the discrepancy between the experimental data and the existing calculation results.

This work is supported by National Basic Research Program of China (Grants No. 2012CB927401, No. 2011CB921902, No. 2013CB921902, No. 2011CB922200), NSFC (Grants No. 91021002, No. 10904090, No. 11174199, No. 11134008), the SCST, China (Grants No. 12JC1405300, No. 13QH1401500, No. 10JC1407100, No. 10PJ1405700, No. 11PJ405200). The Advanced Light Source is supported by the Director, Office of Science, Office of Basic Energy Sciences, of the US Department of Energy under Contract DE-AC02-05CH11231. D.Q. acknowledges additional supports from the Top-notch Young Talents Program.

\end{document}